\documentclass[showpacs,prl,amsmath,amssymb,superscriptaddress,nobibnotes,twocolumn, preprintnumbers]{revtex4-1}
\usepackage{graphicx}
\usepackage{endnotes}
\usepackage{bm}
\usepackage{epsfig}
\usepackage{amsmath}

\newcommand{\ie}{{\it i.e.}} 
\newcommand{\eg}{{\it e.g.}} 
\newcommand{\etal}{{\it et al.}} 
\newcommand{\FSS}{FeSe$_{1-x}$S$_x$}

\begin{document}


\title{Weakening of the diamagnetic shielding in FeSe$_{1-x}$S$_x$ at high pressures}

\author{K.~Y.~Yip}
\author{Y.~C.~Chan}
\author{Q.~Niu}

\affiliation{Department of Physics, The Chinese University of Hong Kong, Shatin, New Territories, Hong Kong, China}

\author{K.~Matsuura}
\author{Y.~Mizukami}
\affiliation{Department of Advanced Materials Science, University of Tokyo, Kashiwa, Chiba 277-8561, Japan}

\author{S.~Kasahara}
\affiliation{Department of Physics, Kyoto University, Sakyo-ku, Kyoto 606-8502, Japan}

\author{Y.~Matsuda}
\affiliation{Department of Physics, Kyoto University, Sakyo-ku, Kyoto 606-8502, Japan}

\author{T.~Shibauchi}
\affiliation{Department of Advanced Materials Science, University of Tokyo, Kashiwa, Chiba 277-8561, Japan}

\author{Swee~K.~Goh}
\email{skgoh@phy.cuhk.edu.hk}
\affiliation{Department of Physics, The Chinese University of Hong Kong, Shatin, New Territories, Hong Kong, China}
\date{\today}


\begin{abstract}
The superconducting transition of FeSe$_{1-x}$S$_x$ with three distinct sulphur concentrations $x$ was studied under hydrostatic pressure up to $\sim$70~kbar via bulk AC susceptibility. The pressure dependence of the superconducting transition temperature ($T_c$) features a small dome-shaped variation at low pressures for $x=0.04$ and $x=0.12$, followed by a more substantial $T_c$ enhancement to a value of around 30~K at moderate pressures. In $x=0.21$, a similar overall pressure dependence of $T_c$ is observed, except that the small dome at low pressures is flattened. For all three concentrations, a significant weakening of the diamagnetic shielding is observed beyond the pressure around which the maximum $T_c$ of 30~K is reached near the verge of pressure-induced magnetic phase. This observation points to a strong competition between the magnetic and high-$T_c$ superconducting states at high pressure in this system. 
\end{abstract}


\maketitle


The importance of nematic order and magnetic order is a central topic in the study of iron-based superconductors \cite{Fernandes2014}. FeSe, although structurally one of the simplest among all known iron-based superconductors, offers a large playground to explore the interplay between superconductivity and other phases \cite{McQueen2009, Kasahara2014, Imai2009, Bendele2010, Terashima2015, Sun2016, Baek2015, Bohmer2015, Hsu2008, Song2011, Mizuguchi2008, Medvedev2009, Braithwaite2009, He2013, Tan2013, Ge2014, Watson2015, Suzuki2015, Kothapalli2016, Wang2016, Terashima2016, Tanatar2016, Kaluarachchi2016}. The superconducting transition temperature ($T_c$), which is 9~K at ambient pressure \cite{Hsu2008, Song2011, Kasahara2014}, can be substantially enhanced either by applying pressure \cite{Mizuguchi2008, Medvedev2009, Braithwaite2009} or by isolating a monolayer of the crystal \cite{He2013, Tan2013, Ge2014}. At $T_s\approx$90~K, FeSe undergoes a tetragonal-to-orthorhombic structural transition \cite{McQueen2009, Baek2015, Bohmer2015}, accompanied by a large electronic anisotropy which breaks $C_4$ rotational symmetry \cite{Song2011, Watson2015, Suzuki2015}. Hence, the transition at $T_s$ is also called a nematic transition. Most importantly, FeSe at ambient pressure is nonmagnetic down to 0~K \cite{Imai2009, Baek2015, Bohmer2015}, while spin density wave (SDW) order can be induced under pressure \cite{Bendele2010, Terashima2015, Kothapalli2016, Wang2016, Terashima2016}, giving rise to an interesting phase diagram featuring nematic, magnetic and superconducting phases \cite{Sun2016}.

Given the richness of the pressure phase diagram, it is a natural attempt to simulate the high pressure effect chemically. A similar approach has been made on BaFe$_2$As$_2$, \eg\ phosphorous has been substituted for arsenic to introduce the chemical pressure effect \cite{Kasahara2010, Goh2010, Klintberg2010}. In FeSe, sulphur has been substituted for selenium, \ie\ FeSe$_{1-x}$S$_x$ \cite{Hosoi2016, Matsuura2017, Xiang2017, Mizuguchi2009, Watson2015b, Moore2015}. Although the lattice constants indeed decrease with increasing $x$, the temperature-$x$ ($T$-$x$) phase diagram is markedly different from the temperature-pressure ($T$-$p$) phase diagram of FeSe \cite{Sun2016, Hosoi2016, Matsuura2017, Xiang2017}. In particular, magnetism is not induced by sulphur. Near $x_c\approx0.17$ where $T_s\rightarrow0$, the nematic susceptibility substantially diverges, suggesting the existence of a nematic quantum critical point \cite{Hosoi2016}. Therefore, the chemical pressure introduced by S substitution is inequivalent to the applied pressure effect. 

The inequivalence between chemical pressure and applied pressure offers the exciting prospect of further fine tuning with a simultaneous application of both tuning parameters. SDW order can also be induced in \FSS\ when physical pressure is applied \cite{Matsuura2017, Xiang2017}. However, at higher $x$, the SDW phase can be completely decoupled from the nematic phase under pressure \cite{Matsuura2017}. Thus, pressure studies of \FSS\ offer an important route to separate the effects of nematic and magnetic fluctuations, allowing a discussion of their relative influence on superconducting pairing.

To study the pressure evolution of superconductivity in \FSS, we have carried out a series of AC susceptibility experiments on single crystals with $x$-values straddling across the nematic quantum critical point. In this manuscript, we report a surprising loss of the AC susceptibility signal associated with the diamagnetic shielding for all crystals studied at moderate pressures, near the maximum $T_c$ reported near the lower border of the pressure-induced magnetic phase, despite the fact that the crystals have been established to be superconducting via high pressure resistivity measurements \cite{Matsuura2017}.


\FSS\ single crystals were synthesized by the chemical vapour transport technique as described elsewhere \cite{Hosoi2016}. The Y-doped Bi$_2$Sr$_2$CaCu$_2$O$_8$ (BSCCO), used as a reference superconductor for some runs, was grown using a method similar to ref. \cite{Eisaki2004}. High pressure AC susceptibility measurements were conducted via a mutual inductance method in moissanite anvil cells \cite{Alireza2003, Goh2010, Klintberg2010}. A modulation coil with $\sim$140 turns was placed around the anvil outside the preindented region of the gasket, and the modulation frequency was between 2~kHz and 14~kHz. A typical pickup coil has $7.5$ turns with a diameter of $\sim$250~$\mu$m and length $\sim$150~$\mu$m, which is enough to accommodate two thin samples and a ruby chip. The skin depth, $\delta$, of a metal is given by $\delta=\sqrt{\rho/(\pi f \mu)}$, where $f$ is the frequency of the measurement, $\rho$ is the resistivity and $\mu$ is the permeability. Assuming $\rho$ of at least 10~$\mu\Omega$cm for \FSS\ \cite{Matsuura2017} and for simplicity, $\mu=\mu_0=4\pi\times10^{-7}$~Hm$^{-1}$, $\delta$ at 14~kHz is calculated to be $\sim1.3$~mm. This value is the lower bound for our estimate and it is larger than all dimensions of the sample. Therefore, the AC field penetrates the sample fully above the superconducting transition, and this is a bulk measurement of the superconducting transition. Glycerin was used as the pressure transmitting fluid. The pressure achieved was determined using ruby fluorescence spectroscopy at room temperature. 

\begin{figure}[!t]\centering
      \resizebox{8.5cm}{!}{
              \includegraphics{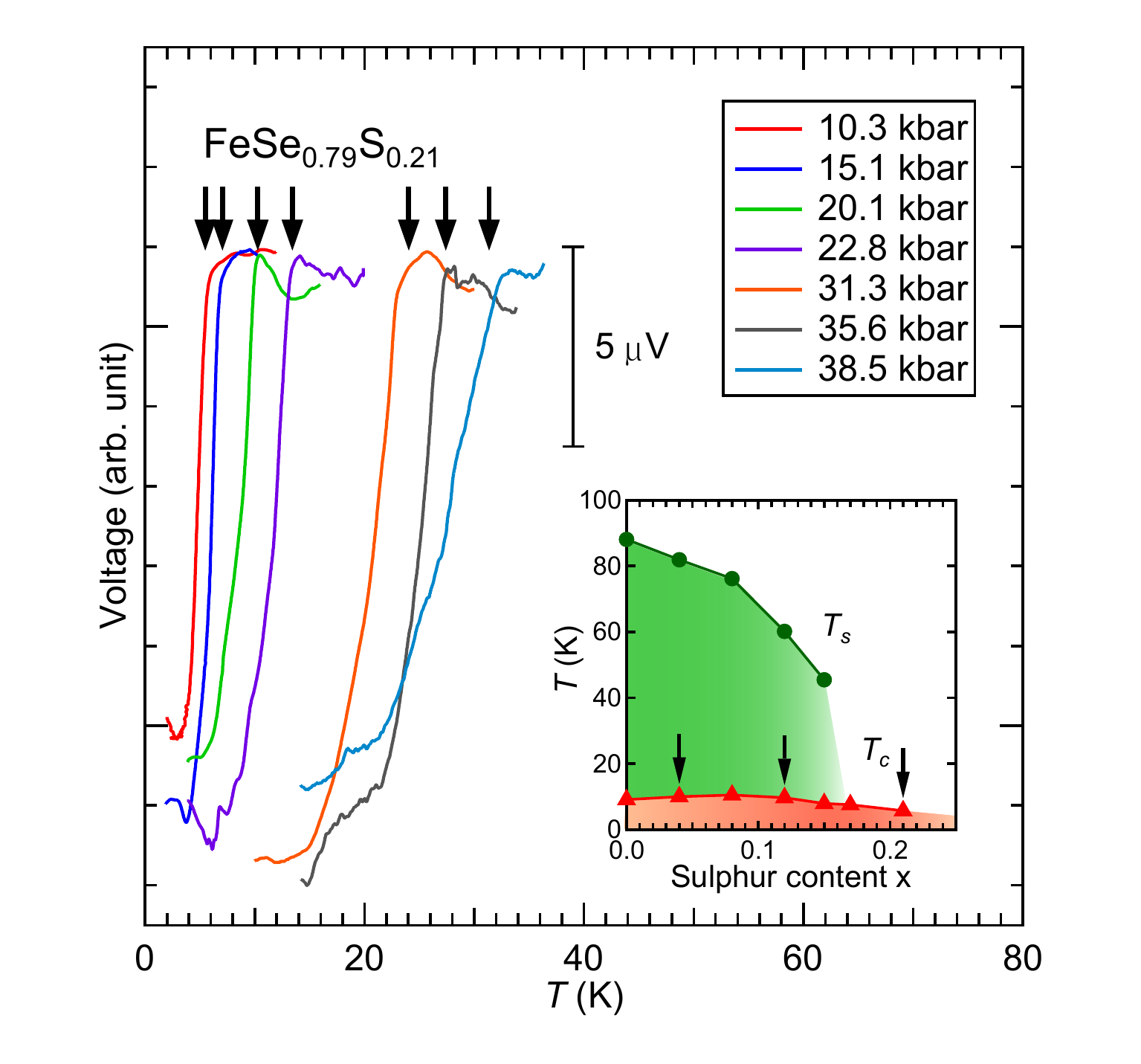}}                				
              \caption{\label{fig1} (Color online) Plot of AC susceptibility against temperature for \FSS\ ($x=0.21$). Voltage drops due to the superconducting transition are shown and the transition temperatures are denoted by arrows. The inset shows the ambient pressure $T$-$x$ phase diagram of \FSS, constructed using data from \cite{Hosoi2016}. The \FSS\ with different S substitutions studied in this work are indicated by arrows ($x=0.04$, $x=0.12$, and $x=0.21$). $T_s$ and $T_c$ denote the structural (nematic) transition temperature and superconducting transition temperature, respectively. 
              }
\end{figure}

\begin{figure}[!t]\centering
       \resizebox{8.5cm}{!}{
              \includegraphics{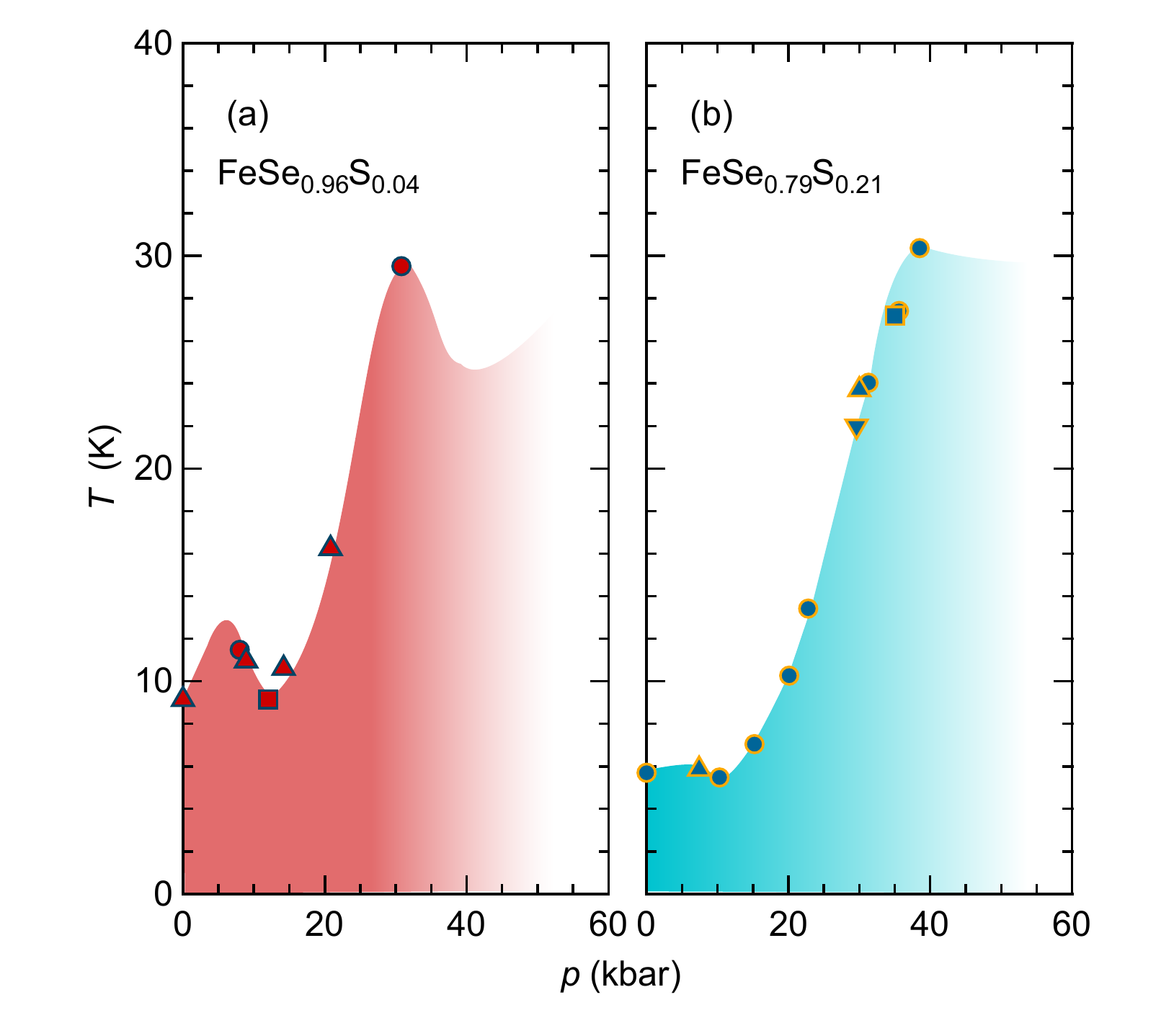}}                				
              \caption{\label{fig2} (Color online) Temperature-pressure phase diagram of \FSS\ with $x=0.04$ (a) and $x=0.21$ (b) as determined by AC susceptibility. Note that data from multiple runs are included using different symbols. For instance, the $T_c$ values extracted from Fig.~\ref{fig1} are shown as solid circles in (b). Collectively, these data give a smooth pressure variation of $T_c$, highlighting excellent experimental reproducibility.
              }
\end{figure}

The previously constructed ambient pressure $T$-$x$ phase diagram of \FSS\ \cite{Hosoi2016} is shown in the inset of Fig.~\ref{fig1}. For this study, we have chosen $x=0.04$, $x=0.12$, and $x=0.21$, covering both the quantum ordered side for the former two compositions and the quantum disordered side for the latter composition, as indicated by the arrows on the $T$-$x$ phase diagram. Our AC susceptibility technique is particularly sensitive to superconducting transitions \cite{Goh2010, Klintberg2010}, with typical signals displayed in the main panel of Fig.~\ref{fig1}. The voltage drop is proportional to the real part of the AC susceptibility, and the arrows in the main panel define the values of $T_c$. The ambient pressure $T_c$ values determined by AC susceptibility for all three concentrations ({\it c.f.} Figs.~\ref{fig2} and \ref{fig4}) are in good agreement with the values measured by resistivity \cite{Matsuura2017, Xiang2017}.

The pressure dependence of $T_c$ is summarised in Figs.~\ref{fig2}a and \ref{fig2}b for $x=0.04$ and $x=0.21$, respectively. Data from multiple runs are included in the construction of the $T$-$p$ phase diagrams. In $x=0.04$, $T_c$ first increases with increasing pressure, then it begins to drop slightly and reaches a local minimum near $\sim$12~kbar.  Similar low pressure behaviour of $T_c(p)$ is observed in $x=0.12$, except that the local minimum is shifted to $\sim$10~kbar. In $x=0.21$, the local minimum is significantly washed out. Recent high-pressure resistivity measurements on \FSS\ up to $x=0.12$ also pointed out the shifting of the local minimum to lower pressure with increasing $x$ \cite{Xiang2017}, which is consistent with our results. The local minimum is thought to be related to the disappearance of the nematic phase when pressure is applied \cite{Hosoi2016}. Therefore, the weakening of this feature is consistent with the fact that $x=0.21$ is outside the nematic phase (inset of Fig.~\ref{fig1}). 
\begin{figure}[!t]\centering
       \resizebox{8.5cm}{!}{
              \includegraphics{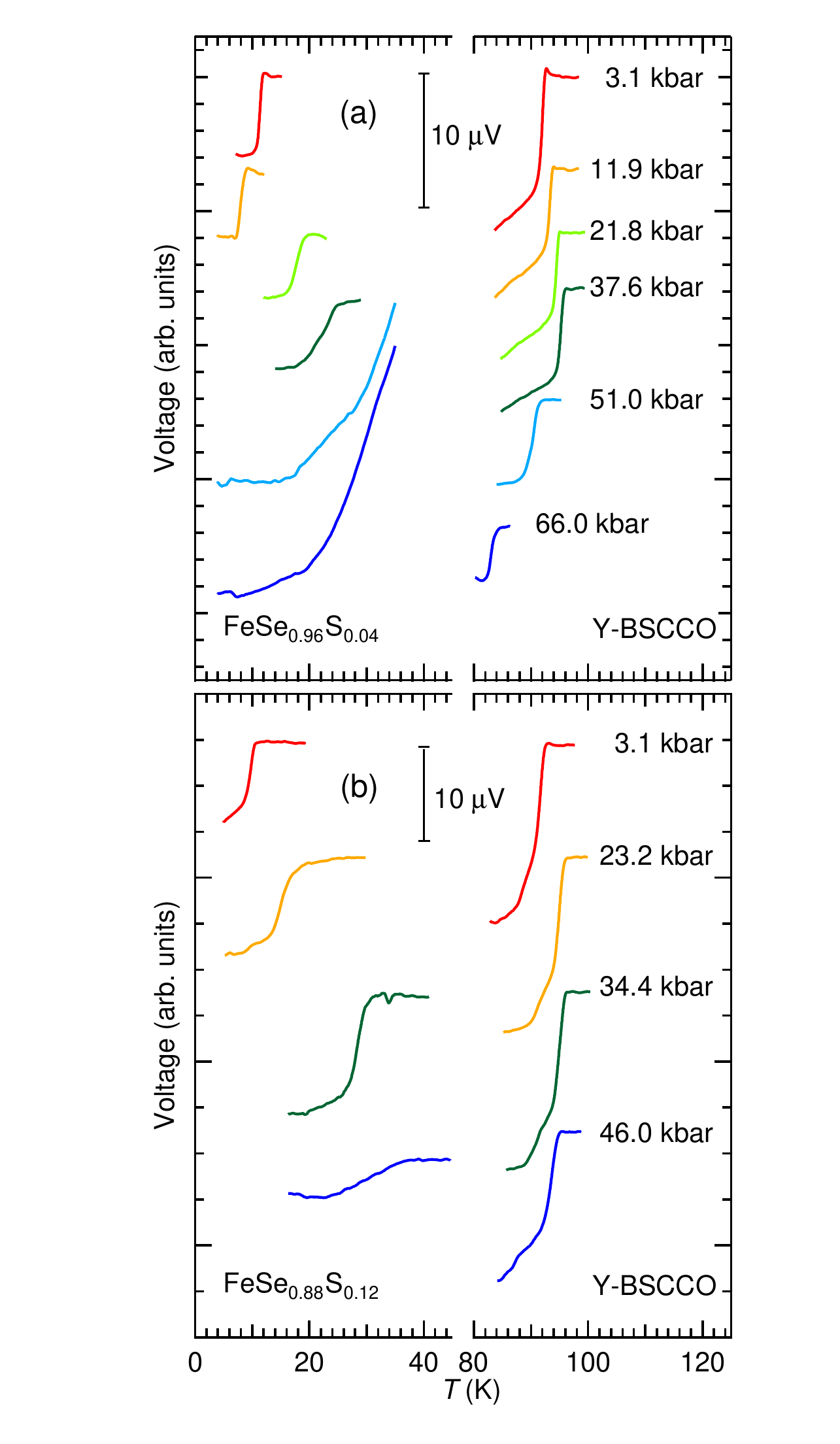}}                				
              \caption{\label{fig3} (Color online) Simultaneous AC susceptibility measurement of (a) Y-BSCCO and $x=0.04$, and (b) Y-BSCCO and $x=0.12$. The measurements were performed in the order of increasing pressure in a single series without changing the sample. The temperature sweeps are offset vertically for clarity. Axis breaks are added to enable a clear display of both the high-temperature and the low-temperature parts, where the superconducting transitions of Y-BSCCO and \FSS\ occur, respectively. The pressure values are shown next to the high-temperature part only.}
\end{figure}

A further increase of pressure leads to a more substantial increase in $T_c$ in all three compositions. In $x=0.04$, $T_c$ reaches $\sim30$~K at $\sim30$~kbar. This is followed by a surprising disappearance of our diamagnetic signal at higher pressure. Similar behaviour is observed in $x=0.12$ and $x=0.21$. 
Representative temperature sweeps showing the disappearance of the diamagnetic signal are presented in Fig. S1 of Supplemental Material.

High pressure resistivity measurements have established the existence of a dome-shaped $T_c(p)$ around this pressure region \cite{Matsuura2017} (see also Fig. S3 of the Supplemental Material). In particular, the crystals of $x=0.12$ used in this study are from the same batch as the crystals used in the resistivity studies of Matsuura \etal\ \cite{Matsuura2017}. Therefore, our AC susceptibility detects only the low pressure ascending part of the $T_c$ dome around this pressure. To rule out instrumentation limitations, \eg\ the malfunctioning of the coils, and to show that the superconducting phase on the higher pressure side of the dome is invisible to AC susceptibility, we design a control experiment by adding a piece of Y-doped BSCCO into our sample space as a benchmark.

Y-doped BSCCO is chosen because of its distinct $T_c$ \cite{Eisaki2004} from \FSS, and hence their superconductivity can both be tracked simultaneously and unambiguously as pressure increases. Fig.~\ref{fig3} displays the results from AC susceptibility measurements with both superconductors inside the same pickup coil. These samples are therefore under identical experimental conditions, such as the pressure environment. At 3.1~kbar, as the temperature of the pressure cell is lowered, a sharp drop in voltage is first detected at around 92~K (Fig.~\ref{fig3}), corresponding to the superconducting transition of Y-BSCCO \cite{Eisaki2004}. Upon further cooling, a second drop in voltage with a comparable magnitude occurs at around 10~K, which is the superconducting transition of \FSS\ with $x=0.04$, consistent with our earlier measurements without Y-BSCCO ({\it c.f.} Fig.~\ref{fig2}a). When pressure is increased, the high-temperature voltage drop is easily identifiable for all pressure values, allowing the construction of $T_c(p)$ for Y-BSCCO. In constrast, the low-temperature voltage drop only remains visible up to a certain pressure. For $x=0.04$, resistivity data shows that $T_c$ ranges between $\sim$20~K and $\sim$30~K between 50~kbar and 70~kbar. As shown in Fig.~\ref{fig3}a, it becomes challenging to identify any clear voltage drop between $\sim$5~K and $\sim$36~K in the temperature sweeps at 51~kbar and 66~kbar. Similar behaviour is observed in $x=0.12$ (Fig.~\ref{fig3}b) and $x=0.21$ (see Supplemental Material Fig.~S2). Therefore, the weakening of the diamagnetic shielding and hence the detection of only the low pressure ascending part of the superconducting dome via AC susceptibility appear to be a generic feature of \FSS.

\begin{figure*}[!t]\centering
       \resizebox{18cm}{!}{
              \includegraphics{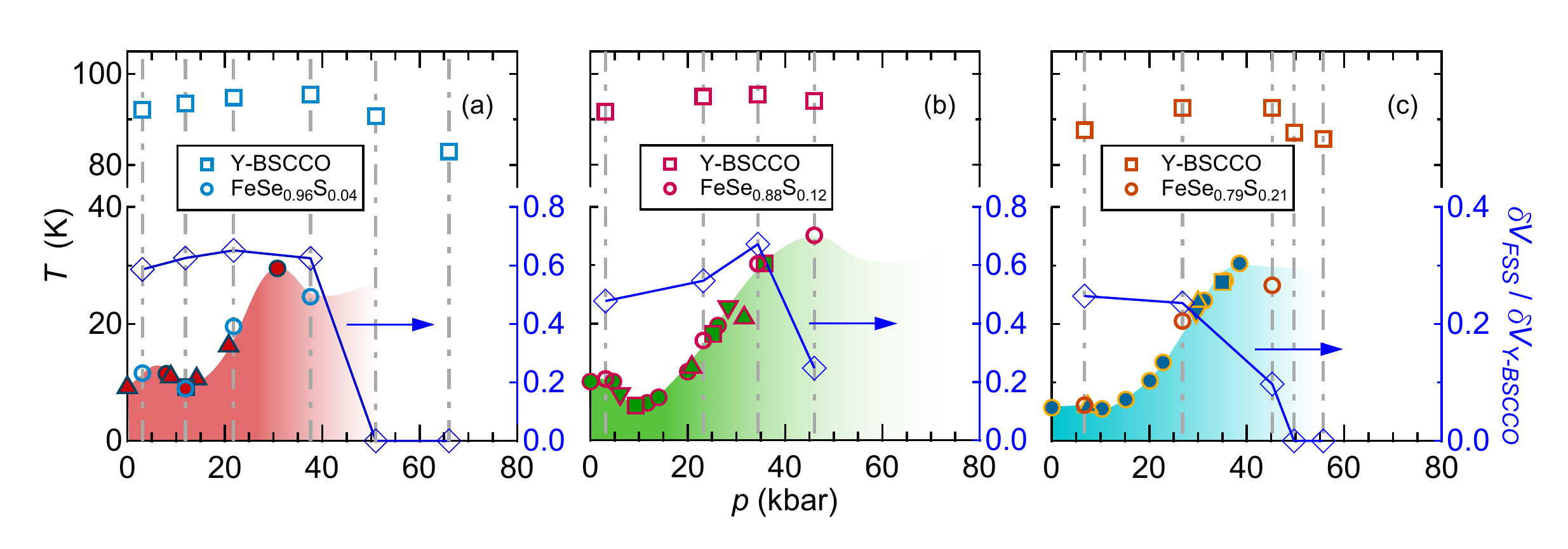}}                				
              \caption{\label{fig4} (Color online) Temperature-pressure phase diagrams, showing the pressure dependence of $T_c$ for (a) $x=0.04$, (b) $x=0.12$, and (c) $x=0.21$. The data from the simultaneous measurements of Y-BSCCO and \FSS\ are denoted by open symbols connected by vertical dashed-dot lines. Note the axis break for the temperature axis. For these measurements, we plot the voltage drop of \FSS\ upon entering the superconducting state, $\delta V_{\rm FSS}$, divided by that of Y-BSCCO, $\delta V_{\rm Y-BSCCO}$ as diamond symbols (right hand axis). For completeness, datapoints from the runs with \FSS\ alone are included as closed symbols. }
              
\end{figure*}
Fig.~\ref{fig4}a summarizes the $T_c(p)$ of Y-BSCCO and $x=0.04$ collected simultaneously using the same pickup coil (open symbols). For comparison, the $T_c(p)$ collected from other runs with $x=0.04$ only is overlaid (solid circles). The overall pressure evolution of $T_c$ in $x=0.04$ is consistent with that determined by resistivity. The $T_c$ of Y-BSCCO experiences a weak pressure dependence, showing a broad superconducting dome with an initial slope d$T_c$/d$p\approx$2~K/GPa. To quantify the disappearance of the diamagnetic shielding further, we calculate the voltage drop upon entering the superconducting state for $x=0.04$, $\delta V_{\rm FSS:0.04}$, divided by that for Y-BSCCO, $\delta V_{\rm Y-BSCCO}$. Assuming Y-BSCCO remains a bulk superconductor under pressure, the ratio $(\delta V_{\rm FSS:0.04}/\delta V_{\rm Y-BSCCO})(p)$ benchmarks the variation of the diamagnetic shielding fraction of $x=0.04$. $(\delta V_{\rm FSS:0.04}/\delta V_{\rm Y-BSCCO})(p)$, shown as diamond symbols in Fig.~\ref{fig4}a is almost pressure independent below $\sim$40~kbar, before plummetting rapidly above this pressure. Figs.~\ref{fig4}b and \ref{fig4}c display the corresponding results for $x=0.12$ and $x=0.21$, respectively. Note that the overall $T_c(p)$ of Y-BSCCO is slightly lower in the pressure cell with $x=0.21$ (Fig.~\ref{fig4}c), presumably due to a slightly different oxygen content. This, however, does not affect the conclusion on the progressive weakening of the diamagnetic shielding in \FSS\ under pressure.


We now discuss the possible reasons for the observed weakening of the diamagnetic shielding under pressure. One possibility could be the extreme sensitivity to pressure inhomogeneity at high pressure. Cubic anvil cells, in which the electrical resistivity data on \FSS\ were collected, are known to provide a highly uniform pressure environment \cite{Mori2004}. If a pressure distribution exists, it would only broaden the transition. Close inspection of the data displayed in Fig.~\ref{fig3}b for $x=0.12$ at 46.0~kbar appears to give a broader transition. However, it is also important to note that the total voltage drop here is much weaker compared with the previous pressure (34.4~kbar). In Y-BSCCO, such a substantial decrease in voltage drop is not observed. Therefore, the reduction in the voltage drop in \FSS\ at high pressures cannot be easily attributed to pressure inhomogeneity.

The fact that the loss of AC susceptibility signals occurs near the verge of magnetism suggests a strong interplay between the two ordered states. It is conceivable that the SDW state competes with superconductivity, gapping out portions of the Fermi surface which would otherwise be available for superconducting pairing. This gives rise to patches of superconducting regions within the sample, thereby reducing the volume fraction of the superconducting component. Recent NMR experiments on FeSe lend support to this scenario \cite{Wang2016}, although some of the previous AC susceptibility measurements at different conditions suggested that the superconductivity in FeSe is bulk in the entire phase diagram \cite{Terashima2015, Braithwaite2009, Sun2016}. We note that the diamagnetic signal in the AC susceptibility may be large even in inhomogeneous superconductors when the intergrain current is strong enough for magnetic shielding. High pressure NMR experiments on \FSS\ will shed light on this issue.

The same AC susceptibility technique has been employed to study the $T_c(p)$ of BaFe$_2$(As$_{1-x}$P$_x$)$_2$ by some of us, following very similar experimental conditions \cite{Klintberg2010}. In the case of BaFe$_2$(As$_{1-x}$P$_x$)$_2$ clear diamagnetic voltage drop can be seen in all crystals studied up to $\sim$70~kbar. For low phosphorous content $x$, the SDW sets in before superconductivity, but no reduction of diamagnetic shielding was detected. However, whether magnetism competes or coexists with superconductivity has been a long-standing issue in the analysis of phase diagrams of many strongly correlated electron systems, including \eg\ CaFe$_2$As$_2$ \cite{Torikachvili2008, Lee2009} and Ce$T$In$_5$ ($T=$ Co, Rh, and Ir) \cite{Petrovic2001, Chen2006}. Therefore, the observed anomalous diamagnetic response in \FSS\ warrants further studies to fully understand the potential impact of the pressure-induced SDW order on the superconducting properties of this system.


In summary, we have tracked the superconducting transition of \FSS\ with $x=0.04$, $0.12$ and $0.21$ up to $\sim$70~kbar using AC susceptibility. The measured $T_c(p)$ shows a local minimum at around 10--15~kbar, which gets progressively washed out with increasing sulphur content. At a slightly higher pressure, a significant loss of the susceptibility signal corresponding to the superconducting transition is observed for all three compositions studied. The weakening of the diamagnetic shielding occurs near the verge of pressure-induced SDW order, which leads to the conclusion that the SDW phase competes with superconductivity, resulting in the reduction of superconducting volume fraction.

\begin{acknowledgments} The authors acknowledge T. Watashige for experimental support. This work was supported by Research Grant Council of Hong Kong (ECS/24300214, GRF/14301316), CUHK Direct Grant (No. 3132719, No. 3132720), CUHK Startup (No. 4930048), and Grant-in-Aids for Scientific Research from Japan Society for the Promotion of Science (No. 25220710,
No. 15H02106, No. 15H03688, No. 16K13837, and No. 15H05852). \end{acknowledgments}
  


\begin{thebibliography}{40}
\expandafter\ifx\csname natexlab\endcsname\relax\def\natexlab#1{#1}\fi
\expandafter\ifx\csname bibnamefont\endcsname\relax
  \def\bibnamefont#1{#1}\fi
\expandafter\ifx\csname bibfnamefont\endcsname\relax
  \def\bibfnamefont#1{#1}\fi
\expandafter\ifx\csname citenamefont\endcsname\relax
  \def\citenamefont#1{#1}\fi
\expandafter\ifx\csname url\endcsname\relax
  \def\url#1{\texttt{#1}}\fi
\expandafter\ifx\csname urlprefix\endcsname\relax\def\urlprefix{URL }\fi
\providecommand{\bibinfo}[2]{#2}
\providecommand{\eprint}[2][]{\url{#2}}

\bibitem[{\citenamefont{Fernandes et~al.}(2014)\citenamefont{Fernandes,
  Chubukov, and Schmalian}}]{Fernandes2014}
\bibinfo{author}{\bibfnamefont{R.}~\bibnamefont{Fernandes}},
  \bibinfo{author}{\bibfnamefont{A.~V.} \bibnamefont{Chubukov}},
  \bibnamefont{and}
  \bibinfo{author}{\bibfnamefont{J.}~\bibnamefont{Schmalian}},
  \bibinfo{journal}{Nat. Phys.} \textbf{\bibinfo{volume}{10}},
  \bibinfo{pages}{97} (\bibinfo{year}{2014}).

\bibitem[{\citenamefont{McQueen et~al.}(2009)\citenamefont{McQueen, Williams,
  Stephens, Tao, Zhu, Ksenofontov, Casper, Felser, and Cava}}]{McQueen2009}
\bibinfo{author}{\bibfnamefont{T.~M.} \bibnamefont{McQueen}},
  \bibinfo{author}{\bibfnamefont{A.~J.} \bibnamefont{Williams}},
  \bibinfo{author}{\bibfnamefont{P.~W.} \bibnamefont{Stephens}},
  \bibinfo{author}{\bibfnamefont{J.}~\bibnamefont{Tao}},
  \bibinfo{author}{\bibfnamefont{Y.}~\bibnamefont{Zhu}},
  \bibinfo{author}{\bibfnamefont{V.}~\bibnamefont{Ksenofontov}},
  \bibinfo{author}{\bibfnamefont{F.}~\bibnamefont{Casper}},
  \bibinfo{author}{\bibfnamefont{C.}~\bibnamefont{Felser}}, \bibnamefont{and}
  \bibinfo{author}{\bibfnamefont{R.~J.} \bibnamefont{Cava}},
  \bibinfo{journal}{Phys. Rev. Lett.} \textbf{\bibinfo{volume}{103}},
  \bibinfo{pages}{057002} (\bibinfo{year}{2009}).

\bibitem[{\citenamefont{Kasahara et~al.}(2014)\citenamefont{Kasahara,
  Watashige, Hanaguri, Kohsaka, Yamashita, Shimoyama, Mizukami, Endo, Ikeda,
  Aoyama et~al.}}]{Kasahara2014}
\bibinfo{author}{\bibfnamefont{S.}~\bibnamefont{Kasahara}},
  \bibinfo{author}{\bibfnamefont{T.}~\bibnamefont{Watashige}},
  \bibinfo{author}{\bibfnamefont{T.}~\bibnamefont{Hanaguri}},
  \bibinfo{author}{\bibfnamefont{Y.}~\bibnamefont{Kohsaka}},
  \bibinfo{author}{\bibfnamefont{T.}~\bibnamefont{Yamashita}},
  \bibinfo{author}{\bibfnamefont{Y.}~\bibnamefont{Shimoyama}},
  \bibinfo{author}{\bibfnamefont{Y.}~\bibnamefont{Mizukami}},
  \bibinfo{author}{\bibfnamefont{R.}~\bibnamefont{Endo}},
  \bibinfo{author}{\bibfnamefont{H.}~\bibnamefont{Ikeda}},
  \bibinfo{author}{\bibfnamefont{K.}~\bibnamefont{Aoyama}},
  \bibnamefont{et~al.}, \bibinfo{journal}{Proc. Natl. Acad. Sci. USA}
  \textbf{\bibinfo{volume}{111}}, \bibinfo{pages}{16309}
  (\bibinfo{year}{2014}).

\bibitem[{\citenamefont{Imai et~al.}(2009)\citenamefont{Imai, Ahilan, Ning,
  McQueen, and Cava}}]{Imai2009}
\bibinfo{author}{\bibfnamefont{T.}~\bibnamefont{Imai}},
  \bibinfo{author}{\bibfnamefont{K.}~\bibnamefont{Ahilan}},
  \bibinfo{author}{\bibfnamefont{F.~L.} \bibnamefont{Ning}},
  \bibinfo{author}{\bibfnamefont{T.~M.} \bibnamefont{McQueen}},
  \bibnamefont{and} \bibinfo{author}{\bibfnamefont{R.~J.} \bibnamefont{Cava}},
  \bibinfo{journal}{Phys. Rev. Lett.} \textbf{\bibinfo{volume}{102}},
  \bibinfo{pages}{177005} (\bibinfo{year}{2009}).

\bibitem[{\citenamefont{Bendele et~al.}(2010)\citenamefont{Bendele, Amato,
  Conder, Elender, Keller, Klauss, Luetkens, Pomjakushina, Raselli, and
  Khasanov}}]{Bendele2010}
\bibinfo{author}{\bibfnamefont{M.}~\bibnamefont{Bendele}},
  \bibinfo{author}{\bibfnamefont{A.}~\bibnamefont{Amato}},
  \bibinfo{author}{\bibfnamefont{K.}~\bibnamefont{Conder}},
  \bibinfo{author}{\bibfnamefont{M.}~\bibnamefont{Elender}},
  \bibinfo{author}{\bibfnamefont{H.}~\bibnamefont{Keller}},
  \bibinfo{author}{\bibfnamefont{H.-H.} \bibnamefont{Klauss}},
  \bibinfo{author}{\bibfnamefont{H.}~\bibnamefont{Luetkens}},
  \bibinfo{author}{\bibfnamefont{E.}~\bibnamefont{Pomjakushina}},
  \bibinfo{author}{\bibfnamefont{A.}~\bibnamefont{Raselli}}, \bibnamefont{and}
  \bibinfo{author}{\bibfnamefont{R.}~\bibnamefont{Khasanov}},
  \bibinfo{journal}{Phys. Rev. Lett.} \textbf{\bibinfo{volume}{104}},
  \bibinfo{pages}{087003} (\bibinfo{year}{2010}).

\bibitem[{\citenamefont{Terashima et~al.}(2015)\citenamefont{Terashima,
  Kikugawa, Kasahara, Watashige, Shibauchi, Matsuda, Wolf, B\"ohmer, Hardy,
  Meingast et~al.}}]{Terashima2015}
\bibinfo{author}{\bibfnamefont{T.}~\bibnamefont{Terashima}},
  \bibinfo{author}{\bibfnamefont{N.}~\bibnamefont{Kikugawa}},
  \bibinfo{author}{\bibfnamefont{S.}~\bibnamefont{Kasahara}},
  \bibinfo{author}{\bibfnamefont{T.}~\bibnamefont{Watashige}},
  \bibinfo{author}{\bibfnamefont{T.}~\bibnamefont{Shibauchi}},
  \bibinfo{author}{\bibfnamefont{Y.}~\bibnamefont{Matsuda}},
  \bibinfo{author}{\bibfnamefont{T.}~\bibnamefont{Wolf}},
  \bibinfo{author}{\bibfnamefont{A.~E.} \bibnamefont{B\"ohmer}},
  \bibinfo{author}{\bibfnamefont{F.}~\bibnamefont{Hardy}},
  \bibinfo{author}{\bibfnamefont{C.}~\bibnamefont{Meingast}},
  \bibnamefont{et~al.}, \bibinfo{journal}{J. Phys. Soc. Jpn.}
  \textbf{\bibinfo{volume}{84}}, \bibinfo{pages}{063701}
  (\bibinfo{year}{2015}).

\bibitem[{\citenamefont{Sun et~al.}(2016)\citenamefont{Sun, Matsuura, Ye,
  Mizukami, Shimozawa, Matsubayashi, Yamashita, Watashige, Kasahara, Matsuda
  et~al.}}]{Sun2016}
\bibinfo{author}{\bibfnamefont{J.~P.} \bibnamefont{Sun}},
  \bibinfo{author}{\bibfnamefont{K.}~\bibnamefont{Matsuura}},
  \bibinfo{author}{\bibfnamefont{G.~Z.} \bibnamefont{Ye}},
  \bibinfo{author}{\bibfnamefont{Y.}~\bibnamefont{Mizukami}},
  \bibinfo{author}{\bibfnamefont{M.}~\bibnamefont{Shimozawa}},
  \bibinfo{author}{\bibfnamefont{K.}~\bibnamefont{Matsubayashi}},
  \bibinfo{author}{\bibfnamefont{M.}~\bibnamefont{Yamashita}},
  \bibinfo{author}{\bibfnamefont{T.}~\bibnamefont{Watashige}},
  \bibinfo{author}{\bibfnamefont{S.}~\bibnamefont{Kasahara}},
  \bibinfo{author}{\bibfnamefont{Y.}~\bibnamefont{Matsuda}},
  \bibnamefont{et~al.}, \bibinfo{journal}{Nat. Commun.}
  \textbf{\bibinfo{volume}{7}}, \bibinfo{pages}{12146} (\bibinfo{year}{2016}).

\bibitem[{\citenamefont{Baek et~al.}(2015)\citenamefont{Baek, Efremov, Ok, Kim,
  van~den Brink, and B�chner}}]{Baek2015}
\bibinfo{author}{\bibfnamefont{S.-H.} \bibnamefont{Baek}},
  \bibinfo{author}{\bibfnamefont{D.~V.} \bibnamefont{Efremov}},
  \bibinfo{author}{\bibfnamefont{J.~M.} \bibnamefont{Ok}},
  \bibinfo{author}{\bibfnamefont{J.~S.} \bibnamefont{Kim}},
  \bibinfo{author}{\bibfnamefont{J.}~\bibnamefont{van~den Brink}},
  \bibnamefont{and}
  \bibinfo{author}{\bibfnamefont{B.}~\bibnamefont{B�chner}},
  \bibinfo{journal}{Nat. Mater.} \textbf{\bibinfo{volume}{14}},
  \bibinfo{pages}{210} (\bibinfo{year}{2015}).

\bibitem[{\citenamefont{B\"{o}hmer et~al.}(2015)\citenamefont{B\"{o}hmer, Arai,
  Hardy, Hattori, Iye, Wolf, L\"{o}hneysen, Ishida, and Meingast}}]{Bohmer2015}
\bibinfo{author}{\bibfnamefont{A.}~\bibnamefont{B\"{o}hmer}},
  \bibinfo{author}{\bibfnamefont{T.}~\bibnamefont{Arai}},
  \bibinfo{author}{\bibfnamefont{F.}~\bibnamefont{Hardy}},
  \bibinfo{author}{\bibfnamefont{T.}~\bibnamefont{Hattori}},
  \bibinfo{author}{\bibfnamefont{T.}~\bibnamefont{Iye}},
  \bibinfo{author}{\bibfnamefont{T.}~\bibnamefont{Wolf}},
  \bibinfo{author}{\bibfnamefont{H.}~\bibnamefont{L\"{o}hneysen}},
  \bibinfo{author}{\bibfnamefont{K.}~\bibnamefont{Ishida}}, \bibnamefont{and}
  \bibinfo{author}{\bibfnamefont{C.}~\bibnamefont{Meingast}},
  \bibinfo{journal}{Phys. Rev. Lett.} \textbf{\bibinfo{volume}{114}},
  \bibinfo{pages}{027001} (\bibinfo{year}{2015}).

\bibitem[{\citenamefont{Hsu et~al.}(2008)\citenamefont{Hsu, Luo, Yeh, Chen,
  Huang, Wu, Lee, Huang, Chu, Yan et~al.}}]{Hsu2008}
\bibinfo{author}{\bibfnamefont{F.-C.} \bibnamefont{Hsu}},
  \bibinfo{author}{\bibfnamefont{J.-Y.} \bibnamefont{Luo}},
  \bibinfo{author}{\bibfnamefont{K.-W.} \bibnamefont{Yeh}},
  \bibinfo{author}{\bibfnamefont{T.-K.} \bibnamefont{Chen}},
  \bibinfo{author}{\bibfnamefont{T.-W.} \bibnamefont{Huang}},
  \bibinfo{author}{\bibfnamefont{P.~M.} \bibnamefont{Wu}},
  \bibinfo{author}{\bibfnamefont{Y.-C.} \bibnamefont{Lee}},
  \bibinfo{author}{\bibfnamefont{Y.-L.} \bibnamefont{Huang}},
  \bibinfo{author}{\bibfnamefont{Y.-Y.} \bibnamefont{Chu}},
  \bibinfo{author}{\bibfnamefont{D.-C.} \bibnamefont{Yan}},
  \bibnamefont{et~al.}, \bibinfo{journal}{Proc. Natl. Acad. Sci. USA}
  \textbf{\bibinfo{volume}{105}}, \bibinfo{pages}{14262}
  (\bibinfo{year}{2008}).

\bibitem[{\citenamefont{Song et~al.}(2011)\citenamefont{Song, Wang, Cheng,
  Jiang, Li, Zhang, Li, He, Wang, Jia et~al.}}]{Song2011}
\bibinfo{author}{\bibfnamefont{C.-L.} \bibnamefont{Song}},
  \bibinfo{author}{\bibfnamefont{Y.-L.} \bibnamefont{Wang}},
  \bibinfo{author}{\bibfnamefont{P.}~\bibnamefont{Cheng}},
  \bibinfo{author}{\bibfnamefont{Y.-P.} \bibnamefont{Jiang}},
  \bibinfo{author}{\bibfnamefont{W.}~\bibnamefont{Li}},
  \bibinfo{author}{\bibfnamefont{T.}~\bibnamefont{Zhang}},
  \bibinfo{author}{\bibfnamefont{Z.}~\bibnamefont{Li}},
  \bibinfo{author}{\bibfnamefont{K.}~\bibnamefont{He}},
  \bibinfo{author}{\bibfnamefont{L.}~\bibnamefont{Wang}},
  \bibinfo{author}{\bibfnamefont{J.-F.} \bibnamefont{Jia}},
  \bibnamefont{et~al.}, \bibinfo{journal}{Science}
  \textbf{\bibinfo{volume}{332}}, \bibinfo{pages}{1410} (\bibinfo{year}{2011}).

\bibitem[{\citenamefont{Mizuguchi et~al.}(2008)\citenamefont{Mizuguchi,
  Tomioka, Tsuda, Yamaguchi, and Takano}}]{Mizuguchi2008}
\bibinfo{author}{\bibfnamefont{Y.}~\bibnamefont{Mizuguchi}},
  \bibinfo{author}{\bibfnamefont{F.}~\bibnamefont{Tomioka}},
  \bibinfo{author}{\bibfnamefont{S.}~\bibnamefont{Tsuda}},
  \bibinfo{author}{\bibfnamefont{T.}~\bibnamefont{Yamaguchi}},
  \bibnamefont{and} \bibinfo{author}{\bibfnamefont{Y.}~\bibnamefont{Takano}},
  \bibinfo{journal}{Appl. Phys. Lett.} \textbf{\bibinfo{volume}{93}},
  \bibinfo{pages}{152505} (\bibinfo{year}{2008}).

\bibitem[{\citenamefont{Medvedev et~al.}(2009)\citenamefont{Medvedev, McQueen,
  Troyan, Palasyuk, Eremets, Cava, Naghavi, Casper, Ksenofontov, Wortmann
  et~al.}}]{Medvedev2009}
\bibinfo{author}{\bibfnamefont{S.}~\bibnamefont{Medvedev}},
  \bibinfo{author}{\bibfnamefont{T.~M.} \bibnamefont{McQueen}},
  \bibinfo{author}{\bibfnamefont{I.~A.} \bibnamefont{Troyan}},
  \bibinfo{author}{\bibfnamefont{T.}~\bibnamefont{Palasyuk}},
  \bibinfo{author}{\bibfnamefont{M.~I.} \bibnamefont{Eremets}},
  \bibinfo{author}{\bibfnamefont{R.~J.} \bibnamefont{Cava}},
  \bibinfo{author}{\bibfnamefont{S.}~\bibnamefont{Naghavi}},
  \bibinfo{author}{\bibfnamefont{F.}~\bibnamefont{Casper}},
  \bibinfo{author}{\bibfnamefont{V.}~\bibnamefont{Ksenofontov}},
  \bibinfo{author}{\bibfnamefont{G.}~\bibnamefont{Wortmann}},
  \bibnamefont{et~al.}, \bibinfo{journal}{Nat. Mater.}
  \textbf{\bibinfo{volume}{8}}, \bibinfo{pages}{630} (\bibinfo{year}{2009}).

\bibitem[{\citenamefont{Braithwaite et~al.}(2009)\citenamefont{Braithwaite,
  Salce, Lapertot, Bourdarot, Marin, Aoki, and Hanfland}}]{Braithwaite2009}
\bibinfo{author}{\bibfnamefont{D.}~\bibnamefont{Braithwaite}},
  \bibinfo{author}{\bibfnamefont{B.}~\bibnamefont{Salce}},
  \bibinfo{author}{\bibfnamefont{G.}~\bibnamefont{Lapertot}},
  \bibinfo{author}{\bibfnamefont{F.}~\bibnamefont{Bourdarot}},
  \bibinfo{author}{\bibfnamefont{C.}~\bibnamefont{Marin}},
  \bibinfo{author}{\bibfnamefont{D.}~\bibnamefont{Aoki}}, \bibnamefont{and}
  \bibinfo{author}{\bibfnamefont{M.}~\bibnamefont{Hanfland}},
  \bibinfo{journal}{J. Phys.: Condens. Matter} \textbf{\bibinfo{volume}{21}},
  \bibinfo{pages}{232202} (\bibinfo{year}{2009}).

\bibitem[{\citenamefont{He et~al.}(2013)\citenamefont{He, He, Zhang, Zhao, Liu,
  Liu, Mou, Ou, Wang, Li et~al.}}]{He2013}
\bibinfo{author}{\bibfnamefont{S.}~\bibnamefont{He}},
  \bibinfo{author}{\bibfnamefont{J.}~\bibnamefont{He}},
  \bibinfo{author}{\bibfnamefont{W.}~\bibnamefont{Zhang}},
  \bibinfo{author}{\bibfnamefont{L.}~\bibnamefont{Zhao}},
  \bibinfo{author}{\bibfnamefont{D.}~\bibnamefont{Liu}},
  \bibinfo{author}{\bibfnamefont{X.}~\bibnamefont{Liu}},
  \bibinfo{author}{\bibfnamefont{D.}~\bibnamefont{Mou}},
  \bibinfo{author}{\bibfnamefont{Y.-B.} \bibnamefont{Ou}},
  \bibinfo{author}{\bibfnamefont{Q.-Y.} \bibnamefont{Wang}},
  \bibinfo{author}{\bibfnamefont{Z.}~\bibnamefont{Li}}, \bibnamefont{et~al.},
  \bibinfo{journal}{Nat. Mater.} \textbf{\bibinfo{volume}{12}},
  \bibinfo{pages}{605} (\bibinfo{year}{2013}).

\bibitem[{\citenamefont{Tan et~al.}(2013)\citenamefont{Tan, Zhang, Xia, Ye,
  Chen, Xie, Peng, Xu, Fan, Xu et~al.}}]{Tan2013}
\bibinfo{author}{\bibfnamefont{S.}~\bibnamefont{Tan}},
  \bibinfo{author}{\bibfnamefont{Y.}~\bibnamefont{Zhang}},
  \bibinfo{author}{\bibfnamefont{M.}~\bibnamefont{Xia}},
  \bibinfo{author}{\bibfnamefont{Z.}~\bibnamefont{Ye}},
  \bibinfo{author}{\bibfnamefont{F.}~\bibnamefont{Chen}},
  \bibinfo{author}{\bibfnamefont{X.}~\bibnamefont{Xie}},
  \bibinfo{author}{\bibfnamefont{R.}~\bibnamefont{Peng}},
  \bibinfo{author}{\bibfnamefont{D.}~\bibnamefont{Xu}},
  \bibinfo{author}{\bibfnamefont{Q.}~\bibnamefont{Fan}},
  \bibinfo{author}{\bibfnamefont{H.}~\bibnamefont{Xu}}, \bibnamefont{et~al.},
  \bibinfo{journal}{Nat. Mater.} \textbf{\bibinfo{volume}{12}},
  \bibinfo{pages}{634} (\bibinfo{year}{2013}).

\bibitem[{\citenamefont{Ge et~al.}(2014)\citenamefont{Ge, Liu, Liu, Gao, Qian,
  Xue, Liu, and Jia}}]{Ge2014}
\bibinfo{author}{\bibfnamefont{J.-F.} \bibnamefont{Ge}},
  \bibinfo{author}{\bibfnamefont{Z.-L.} \bibnamefont{Liu}},
  \bibinfo{author}{\bibfnamefont{C.}~\bibnamefont{Liu}},
  \bibinfo{author}{\bibfnamefont{C.-L.} \bibnamefont{Gao}},
  \bibinfo{author}{\bibfnamefont{D.}~\bibnamefont{Qian}},
  \bibinfo{author}{\bibfnamefont{Q.-K.} \bibnamefont{Xue}},
  \bibinfo{author}{\bibfnamefont{Y.}~\bibnamefont{Liu}}, \bibnamefont{and}
  \bibinfo{author}{\bibfnamefont{J.-F.} \bibnamefont{Jia}},
  \bibinfo{journal}{Nat. Mater.} \textbf{\bibinfo{volume}{14}},
  \bibinfo{pages}{285} (\bibinfo{year}{2014}).

\bibitem[{\citenamefont{Watson et~al.}(2015{\natexlab{a}})\citenamefont{Watson,
  Kim, Haghighirad, Davies, McCollam, Narayanan, Blake, Chen, Ghannadzadeh,
  Schofield et~al.}}]{Watson2015}
\bibinfo{author}{\bibfnamefont{M.~D.} \bibnamefont{Watson}},
  \bibinfo{author}{\bibfnamefont{T.~K.} \bibnamefont{Kim}},
  \bibinfo{author}{\bibfnamefont{A.~A.} \bibnamefont{Haghighirad}},
  \bibinfo{author}{\bibfnamefont{N.~R.} \bibnamefont{Davies}},
  \bibinfo{author}{\bibfnamefont{A.}~\bibnamefont{McCollam}},
  \bibinfo{author}{\bibfnamefont{A.}~\bibnamefont{Narayanan}},
  \bibinfo{author}{\bibfnamefont{S.~F.} \bibnamefont{Blake}},
  \bibinfo{author}{\bibfnamefont{Y.~L.} \bibnamefont{Chen}},
  \bibinfo{author}{\bibfnamefont{S.}~\bibnamefont{Ghannadzadeh}},
  \bibinfo{author}{\bibfnamefont{A.~J.} \bibnamefont{Schofield}},
  \bibnamefont{et~al.}, \bibinfo{journal}{Phys. Rev. B}
  \textbf{\bibinfo{volume}{91}}, \bibinfo{pages}{155106}
  (\bibinfo{year}{2015}{\natexlab{a}}).

\bibitem[{\citenamefont{Suzuki et~al.}(2015)\citenamefont{Suzuki, Shimojima,
  Sonobe, Nakamura, Sakano, Tsuji, Omachi, Yoshioka, Kuwata-Gonokami, Watashige
  et~al.}}]{Suzuki2015}
\bibinfo{author}{\bibfnamefont{Y.}~\bibnamefont{Suzuki}},
  \bibinfo{author}{\bibfnamefont{T.}~\bibnamefont{Shimojima}},
  \bibinfo{author}{\bibfnamefont{T.}~\bibnamefont{Sonobe}},
  \bibinfo{author}{\bibfnamefont{A.}~\bibnamefont{Nakamura}},
  \bibinfo{author}{\bibfnamefont{M.}~\bibnamefont{Sakano}},
  \bibinfo{author}{\bibfnamefont{H.}~\bibnamefont{Tsuji}},
  \bibinfo{author}{\bibfnamefont{J.}~\bibnamefont{Omachi}},
  \bibinfo{author}{\bibfnamefont{K.}~\bibnamefont{Yoshioka}},
  \bibinfo{author}{\bibfnamefont{M.}~\bibnamefont{Kuwata-Gonokami}},
  \bibinfo{author}{\bibfnamefont{T.}~\bibnamefont{Watashige}},
  \bibnamefont{et~al.}, \bibinfo{journal}{Phys. Rev. B}
  \textbf{\bibinfo{volume}{92}}, \bibinfo{pages}{205117}
  (\bibinfo{year}{2015}).

\bibitem[{\citenamefont{Kothapalli et~al.}(2016)\citenamefont{Kothapalli,
  B\"{o}hmer, Jayasekara, Ueland, Das, Sapkota, Taufour, Xiao, Alp, Bud'ko
  et~al.}}]{Kothapalli2016}
\bibinfo{author}{\bibfnamefont{K.}~\bibnamefont{Kothapalli}},
  \bibinfo{author}{\bibfnamefont{A.~E.} \bibnamefont{B\"{o}hmer}},
  \bibinfo{author}{\bibfnamefont{W.~T.} \bibnamefont{Jayasekara}},
  \bibinfo{author}{\bibfnamefont{B.~G.} \bibnamefont{Ueland}},
  \bibinfo{author}{\bibfnamefont{P.}~\bibnamefont{Das}},
  \bibinfo{author}{\bibfnamefont{A.}~\bibnamefont{Sapkota}},
  \bibinfo{author}{\bibfnamefont{V.}~\bibnamefont{Taufour}},
  \bibinfo{author}{\bibfnamefont{Y.}~\bibnamefont{Xiao}},
  \bibinfo{author}{\bibfnamefont{E.}~\bibnamefont{Alp}},
  \bibinfo{author}{\bibfnamefont{S.~L.} \bibnamefont{Bud'ko}},
  \bibnamefont{et~al.}, \bibinfo{journal}{Nat. Commun.}
  \textbf{\bibinfo{volume}{7}}, \bibinfo{pages}{12728} (\bibinfo{year}{2016}).

\bibitem[{\citenamefont{Wang et~al.}(2016)\citenamefont{Wang, Sun, Cui, Song,
  Li, Yu, Lei, and Yu}}]{Wang2016}
\bibinfo{author}{\bibfnamefont{P.~S.} \bibnamefont{Wang}},
  \bibinfo{author}{\bibfnamefont{S.~S.} \bibnamefont{Sun}},
  \bibinfo{author}{\bibfnamefont{Y.}~\bibnamefont{Cui}},
  \bibinfo{author}{\bibfnamefont{W.~H.} \bibnamefont{Song}},
  \bibinfo{author}{\bibfnamefont{T.~R.} \bibnamefont{Li}},
  \bibinfo{author}{\bibfnamefont{R.}~\bibnamefont{Yu}},
  \bibinfo{author}{\bibfnamefont{H.}~\bibnamefont{Lei}}, \bibnamefont{and}
  \bibinfo{author}{\bibfnamefont{W.}~\bibnamefont{Yu}}, \bibinfo{journal}{Phys.
  Rev. Lett.} \textbf{\bibinfo{volume}{117}}, \bibinfo{pages}{237001}
  (\bibinfo{year}{2016}).

\bibitem[{\citenamefont{Terashima et~al.}(2016)\citenamefont{Terashima,
  Kikugawa, Kasahara, Watashige, Matsuda, Shibauchi, and Uji}}]{Terashima2016}
\bibinfo{author}{\bibfnamefont{T.}~\bibnamefont{Terashima}},
  \bibinfo{author}{\bibfnamefont{N.}~\bibnamefont{Kikugawa}},
  \bibinfo{author}{\bibfnamefont{S.}~\bibnamefont{Kasahara}},
  \bibinfo{author}{\bibfnamefont{T.}~\bibnamefont{Watashige}},
  \bibinfo{author}{\bibfnamefont{Y.}~\bibnamefont{Matsuda}},
  \bibinfo{author}{\bibfnamefont{T.}~\bibnamefont{Shibauchi}},
  \bibnamefont{and} \bibinfo{author}{\bibfnamefont{S.}~\bibnamefont{Uji}},
  \bibinfo{journal}{Phys. Rev. B} \textbf{\bibinfo{volume}{93}},
  \bibinfo{pages}{180503} (\bibinfo{year}{2016}).

\bibitem[{\citenamefont{Tanatar et~al.}(2016)\citenamefont{Tanatar, B\"ohmer,
  Timmons, Sch\"utt, Drachuck, Taufour, Kothapalli, Kreyssig, Bud'ko, Canfield
  et~al.}}]{Tanatar2016}
\bibinfo{author}{\bibfnamefont{M.~A.} \bibnamefont{Tanatar}},
  \bibinfo{author}{\bibfnamefont{A.~E.} \bibnamefont{B\"ohmer}},
  \bibinfo{author}{\bibfnamefont{E.~I.} \bibnamefont{Timmons}},
  \bibinfo{author}{\bibfnamefont{M.}~\bibnamefont{Sch\"utt}},
  \bibinfo{author}{\bibfnamefont{G.}~\bibnamefont{Drachuck}},
  \bibinfo{author}{\bibfnamefont{V.}~\bibnamefont{Taufour}},
  \bibinfo{author}{\bibfnamefont{K.}~\bibnamefont{Kothapalli}},
  \bibinfo{author}{\bibfnamefont{A.}~\bibnamefont{Kreyssig}},
  \bibinfo{author}{\bibfnamefont{S.~L.} \bibnamefont{Bud'ko}},
  \bibinfo{author}{\bibfnamefont{P.~C.} \bibnamefont{Canfield}},
  \bibnamefont{et~al.}, \bibinfo{journal}{Phys. Rev. Lett.}
  \textbf{\bibinfo{volume}{117}}, \bibinfo{pages}{127001}
  (\bibinfo{year}{2016}).

\bibitem[{\citenamefont{Kaluarachchi et~al.}(2016)\citenamefont{Kaluarachchi,
  Taufour, B\"ohmer, Tanatar, Bud'ko, Kogan, Prozorov, and
  Canfield}}]{Kaluarachchi2016}
\bibinfo{author}{\bibfnamefont{U.~S.} \bibnamefont{Kaluarachchi}},
  \bibinfo{author}{\bibfnamefont{V.}~\bibnamefont{Taufour}},
  \bibinfo{author}{\bibfnamefont{A.~E.} \bibnamefont{B\"ohmer}},
  \bibinfo{author}{\bibfnamefont{M.~A.} \bibnamefont{Tanatar}},
  \bibinfo{author}{\bibfnamefont{S.~L.} \bibnamefont{Bud'ko}},
  \bibinfo{author}{\bibfnamefont{V.~G.} \bibnamefont{Kogan}},
  \bibinfo{author}{\bibfnamefont{R.}~\bibnamefont{Prozorov}}, \bibnamefont{and}
  \bibinfo{author}{\bibfnamefont{P.~C.} \bibnamefont{Canfield}},
  \bibinfo{journal}{Phys. Rev. B} \textbf{\bibinfo{volume}{93}},
  \bibinfo{pages}{064503} (\bibinfo{year}{2016}).

\bibitem[{\citenamefont{Kasahara et~al.}(2010)\citenamefont{Kasahara,
  Shibauchi, Hashimoto, Ikada, Tonegawa, Okazaki, Shishido, Ikeda, Takeya,
  Hirata et~al.}}]{Kasahara2010}
\bibinfo{author}{\bibfnamefont{S.}~\bibnamefont{Kasahara}},
  \bibinfo{author}{\bibfnamefont{T.}~\bibnamefont{Shibauchi}},
  \bibinfo{author}{\bibfnamefont{K.}~\bibnamefont{Hashimoto}},
  \bibinfo{author}{\bibfnamefont{K.}~\bibnamefont{Ikada}},
  \bibinfo{author}{\bibfnamefont{S.}~\bibnamefont{Tonegawa}},
  \bibinfo{author}{\bibfnamefont{R.}~\bibnamefont{Okazaki}},
  \bibinfo{author}{\bibfnamefont{H.}~\bibnamefont{Shishido}},
  \bibinfo{author}{\bibfnamefont{H.}~\bibnamefont{Ikeda}},
  \bibinfo{author}{\bibfnamefont{H.}~\bibnamefont{Takeya}},
  \bibinfo{author}{\bibfnamefont{K.}~\bibnamefont{Hirata}},
  \bibnamefont{et~al.}, \bibinfo{journal}{Phys. Rev. B}
  \textbf{\bibinfo{volume}{81}}, \bibinfo{pages}{184519}
  (\bibinfo{year}{2010}).

\bibitem[{\citenamefont{Goh et~al.}(2010)\citenamefont{Goh, Nakai, Ishida,
  Klintberg, Ihara, Kasahara, Shibauchi, Matsuda, and Terashima}}]{Goh2010}
\bibinfo{author}{\bibfnamefont{S.~K.} \bibnamefont{Goh}},
  \bibinfo{author}{\bibfnamefont{Y.}~\bibnamefont{Nakai}},
  \bibinfo{author}{\bibfnamefont{K.}~\bibnamefont{Ishida}},
  \bibinfo{author}{\bibfnamefont{L.~E.} \bibnamefont{Klintberg}},
  \bibinfo{author}{\bibfnamefont{Y.}~\bibnamefont{Ihara}},
  \bibinfo{author}{\bibfnamefont{S.}~\bibnamefont{Kasahara}},
  \bibinfo{author}{\bibfnamefont{T.}~\bibnamefont{Shibauchi}},
  \bibinfo{author}{\bibfnamefont{Y.}~\bibnamefont{Matsuda}}, \bibnamefont{and}
  \bibinfo{author}{\bibfnamefont{T.}~\bibnamefont{Terashima}},
  \bibinfo{journal}{Phys. Rev. B} \textbf{\bibinfo{volume}{82}},
  \bibinfo{pages}{094502} (\bibinfo{year}{2010}).

\bibitem[{\citenamefont{Klintberg et~al.}(2010)\citenamefont{Klintberg, Goh,
  Kasahara, Nakai, Ishida, Sutherland, Shibauchi, Matsuda, and
  Terashima}}]{Klintberg2010}
\bibinfo{author}{\bibfnamefont{L.~E.} \bibnamefont{Klintberg}},
  \bibinfo{author}{\bibfnamefont{S.~K.} \bibnamefont{Goh}},
  \bibinfo{author}{\bibfnamefont{S.}~\bibnamefont{Kasahara}},
  \bibinfo{author}{\bibfnamefont{Y.}~\bibnamefont{Nakai}},
  \bibinfo{author}{\bibfnamefont{K.}~\bibnamefont{Ishida}},
  \bibinfo{author}{\bibfnamefont{M.}~\bibnamefont{Sutherland}},
  \bibinfo{author}{\bibfnamefont{T.}~\bibnamefont{Shibauchi}},
  \bibinfo{author}{\bibfnamefont{Y.}~\bibnamefont{Matsuda}}, \bibnamefont{and}
  \bibinfo{author}{\bibfnamefont{T.}~\bibnamefont{Terashima}},
  \bibinfo{journal}{J. Phys. Soc. Jpn.} \textbf{\bibinfo{volume}{79}},
  \bibinfo{pages}{123706} (\bibinfo{year}{2010}).

\bibitem[{\citenamefont{Hosoi et~al.}(2016)\citenamefont{Hosoi, Matsuura,
  Ishida, Wang, Mizukami, Watashige, Kasahara, Matsuda, and
  Shibauchi}}]{Hosoi2016}
\bibinfo{author}{\bibfnamefont{S.}~\bibnamefont{Hosoi}},
  \bibinfo{author}{\bibfnamefont{K.}~\bibnamefont{Matsuura}},
  \bibinfo{author}{\bibfnamefont{K.}~\bibnamefont{Ishida}},
  \bibinfo{author}{\bibfnamefont{H.}~\bibnamefont{Wang}},
  \bibinfo{author}{\bibfnamefont{Y.}~\bibnamefont{Mizukami}},
  \bibinfo{author}{\bibfnamefont{T.}~\bibnamefont{Watashige}},
  \bibinfo{author}{\bibfnamefont{S.}~\bibnamefont{Kasahara}},
  \bibinfo{author}{\bibfnamefont{Y.}~\bibnamefont{Matsuda}}, \bibnamefont{and}
  \bibinfo{author}{\bibfnamefont{T.}~\bibnamefont{Shibauchi}},
  \bibinfo{journal}{Proc. Natl. Acad. Sci. USA} \textbf{\bibinfo{volume}{113}},
  \bibinfo{pages}{8139} (\bibinfo{year}{2016}).

\bibitem[{\citenamefont{{Matsuura} et~al.}(2017)\citenamefont{{Matsuura},
  {Mizukami}, {Arai}, {Sugimura}, {Maejima}, {Machida}, {Watanuki}, {Fukuda},
  {Yajima}, {Hiroi} et~al.}}]{Matsuura2017}
\bibinfo{author}{\bibfnamefont{K.}~\bibnamefont{{Matsuura}}},
  \bibinfo{author}{\bibfnamefont{Y.}~\bibnamefont{{Mizukami}}},
  \bibinfo{author}{\bibfnamefont{Y.}~\bibnamefont{{Arai}}},
  \bibinfo{author}{\bibfnamefont{Y.}~\bibnamefont{{Sugimura}}},
  \bibinfo{author}{\bibfnamefont{N.}~\bibnamefont{{Maejima}}},
  \bibinfo{author}{\bibfnamefont{A.}~\bibnamefont{{Machida}}},
  \bibinfo{author}{\bibfnamefont{T.}~\bibnamefont{{Watanuki}}},
  \bibinfo{author}{\bibfnamefont{T.}~\bibnamefont{{Fukuda}}},
  \bibinfo{author}{\bibfnamefont{T.}~\bibnamefont{{Yajima}}},
  \bibinfo{author}{\bibfnamefont{Z.}~\bibnamefont{{Hiroi}}},
  \bibnamefont{et~al.}, \bibinfo{journal}{arXiv e-prints}
  (\bibinfo{year}{2017}), \eprint{1704.02057}.

\bibitem[{\citenamefont{{Xiang} et~al.}(2017)\citenamefont{{Xiang},
  {Kaluarachchi}, {B{\"o}hmer}, {Taufour}, {Tanatar}, {Prozorov}, {Bud'ko}, and
  {Canfield}}}]{Xiang2017}
\bibinfo{author}{\bibfnamefont{L.}~\bibnamefont{{Xiang}}},
  \bibinfo{author}{\bibfnamefont{U.~S.} \bibnamefont{{Kaluarachchi}}},
  \bibinfo{author}{\bibfnamefont{A.~E.} \bibnamefont{{B{\"o}hmer}}},
  \bibinfo{author}{\bibfnamefont{V.}~\bibnamefont{{Taufour}}},
  \bibinfo{author}{\bibfnamefont{M.~A.} \bibnamefont{{Tanatar}}},
  \bibinfo{author}{\bibfnamefont{R.}~\bibnamefont{{Prozorov}}},
  \bibinfo{author}{\bibfnamefont{S.~L.} \bibnamefont{{Bud'ko}}},
  \bibnamefont{and} \bibinfo{author}{\bibfnamefont{P.~C.}
  \bibnamefont{{Canfield}}}, \bibinfo{journal}{arXiv e-prints}
  (\bibinfo{year}{2017}), \eprint{1704.04999}.

\bibitem[{\citenamefont{Mizuguchi et~al.}(2009)\citenamefont{Mizuguchi,
  Tomioka, Tsuda, Yamaguchi, and Takano}}]{Mizuguchi2009}
\bibinfo{author}{\bibfnamefont{Y.}~\bibnamefont{Mizuguchi}},
  \bibinfo{author}{\bibfnamefont{F.}~\bibnamefont{Tomioka}},
  \bibinfo{author}{\bibfnamefont{S.}~\bibnamefont{Tsuda}},
  \bibinfo{author}{\bibfnamefont{T.}~\bibnamefont{Yamaguchi}},
  \bibnamefont{and} \bibinfo{author}{\bibfnamefont{Y.}~\bibnamefont{Takano}},
  \bibinfo{journal}{J. Phys. Soc. Jpn.} \textbf{\bibinfo{volume}{78}},
  \bibinfo{pages}{074712} (\bibinfo{year}{2009}).

\bibitem[{\citenamefont{Watson et~al.}(2015{\natexlab{b}})\citenamefont{Watson,
  Kim, Haghighirad, Blake, Davies, Hoesch, Wolf, and Coldea}}]{Watson2015b}
\bibinfo{author}{\bibfnamefont{M.~D.} \bibnamefont{Watson}},
  \bibinfo{author}{\bibfnamefont{T.~K.} \bibnamefont{Kim}},
  \bibinfo{author}{\bibfnamefont{A.~A.} \bibnamefont{Haghighirad}},
  \bibinfo{author}{\bibfnamefont{S.~F.} \bibnamefont{Blake}},
  \bibinfo{author}{\bibfnamefont{N.~R.} \bibnamefont{Davies}},
  \bibinfo{author}{\bibfnamefont{M.}~\bibnamefont{Hoesch}},
  \bibinfo{author}{\bibfnamefont{T.}~\bibnamefont{Wolf}}, \bibnamefont{and}
  \bibinfo{author}{\bibfnamefont{A.~I.} \bibnamefont{Coldea}},
  \bibinfo{journal}{Phys. Rev. B} \textbf{\bibinfo{volume}{92}},
  \bibinfo{pages}{121108} (\bibinfo{year}{2015}{\natexlab{b}}).

\bibitem[{\citenamefont{Moore et~al.}(2015)\citenamefont{Moore, Curtis,
  Giorgio, Lechner, Abdel-Hafiez, Volkova, Vasiliev, Chareev, Karapetrov, and
  Iavarone}}]{Moore2015}
\bibinfo{author}{\bibfnamefont{S.~A.} \bibnamefont{Moore}},
  \bibinfo{author}{\bibfnamefont{J.~L.} \bibnamefont{Curtis}},
  \bibinfo{author}{\bibfnamefont{C.~D.} \bibnamefont{Giorgio}},
  \bibinfo{author}{\bibfnamefont{E.}~\bibnamefont{Lechner}},
  \bibinfo{author}{\bibfnamefont{M.}~\bibnamefont{Abdel-Hafiez}},
  \bibinfo{author}{\bibfnamefont{O.~S.} \bibnamefont{Volkova}},
  \bibinfo{author}{\bibfnamefont{A.~N.} \bibnamefont{Vasiliev}},
  \bibinfo{author}{\bibfnamefont{D.~A.} \bibnamefont{Chareev}},
  \bibinfo{author}{\bibfnamefont{G.}~\bibnamefont{Karapetrov}},
  \bibnamefont{and} \bibinfo{author}{\bibfnamefont{M.}~\bibnamefont{Iavarone}},
  \bibinfo{journal}{Phys. Rev. B} \textbf{\bibinfo{volume}{92}},
  \bibinfo{pages}{235113} (\bibinfo{year}{2015}).

\bibitem[{\citenamefont{Eisaki et~al.}(2004)\citenamefont{Eisaki, Kaneko, Feng,
  Damascelli, Mang, Shen, Shen, and Greven}}]{Eisaki2004}
\bibinfo{author}{\bibfnamefont{H.}~\bibnamefont{Eisaki}},
  \bibinfo{author}{\bibfnamefont{N.}~\bibnamefont{Kaneko}},
  \bibinfo{author}{\bibfnamefont{D.~L.} \bibnamefont{Feng}},
  \bibinfo{author}{\bibfnamefont{A.}~\bibnamefont{Damascelli}},
  \bibinfo{author}{\bibfnamefont{P.~K.} \bibnamefont{Mang}},
  \bibinfo{author}{\bibfnamefont{K.~M.} \bibnamefont{Shen}},
  \bibinfo{author}{\bibfnamefont{Z.-X.} \bibnamefont{Shen}}, \bibnamefont{and}
  \bibinfo{author}{\bibfnamefont{M.}~\bibnamefont{Greven}},
  \bibinfo{journal}{Phys. Rev. B} \textbf{\bibinfo{volume}{69}},
  \bibinfo{pages}{064512} (\bibinfo{year}{2004}).

\bibitem[{\citenamefont{Alireza and Julian}(2003)}]{Alireza2003}
\bibinfo{author}{\bibfnamefont{P.~L.} \bibnamefont{Alireza}} \bibnamefont{and}
  \bibinfo{author}{\bibfnamefont{S.~R.} \bibnamefont{Julian}},
  \bibinfo{journal}{Rev. Sci. Instrum.} \textbf{\bibinfo{volume}{74}},
  \bibinfo{pages}{4728} (\bibinfo{year}{2003}).

\bibitem[{\citenamefont{Mori et~al.}(2004)\citenamefont{Mori, Takahashi, and
  Takeshita}}]{Mori2004}
\bibinfo{author}{\bibfnamefont{N.}~\bibnamefont{Mori}},
  \bibinfo{author}{\bibfnamefont{H.}~\bibnamefont{Takahashi}},
  \bibnamefont{and}
  \bibinfo{author}{\bibfnamefont{N.}~\bibnamefont{Takeshita}},
  \bibinfo{journal}{High Press. Res.} \textbf{\bibinfo{volume}{24}},
  \bibinfo{pages}{225} (\bibinfo{year}{2004}).

\bibitem[{\citenamefont{Torikachvili et~al.}(2008)\citenamefont{Torikachvili,
  Bud'ko, Ni, and Canfield}}]{Torikachvili2008}
\bibinfo{author}{\bibfnamefont{M.~S.} \bibnamefont{Torikachvili}},
  \bibinfo{author}{\bibfnamefont{S.~L.} \bibnamefont{Bud'ko}},
  \bibinfo{author}{\bibfnamefont{N.}~\bibnamefont{Ni}}, \bibnamefont{and}
  \bibinfo{author}{\bibfnamefont{P.~C.} \bibnamefont{Canfield}},
  \bibinfo{journal}{Phys. Rev. Lett.} \textbf{\bibinfo{volume}{101}},
  \bibinfo{pages}{057006} (\bibinfo{year}{2008}).

\bibitem[{\citenamefont{Lee et~al.}(2009)\citenamefont{Lee, Park, Park,
  Sidorov, Ronning, Bauer, and Thompson}}]{Lee2009}
\bibinfo{author}{\bibfnamefont{H.}~\bibnamefont{Lee}},
  \bibinfo{author}{\bibfnamefont{E.}~\bibnamefont{Park}},
  \bibinfo{author}{\bibfnamefont{T.}~\bibnamefont{Park}},
  \bibinfo{author}{\bibfnamefont{V.~A.} \bibnamefont{Sidorov}},
  \bibinfo{author}{\bibfnamefont{F.}~\bibnamefont{Ronning}},
  \bibinfo{author}{\bibfnamefont{E.~D.} \bibnamefont{Bauer}}, \bibnamefont{and}
  \bibinfo{author}{\bibfnamefont{J.~D.} \bibnamefont{Thompson}},
  \bibinfo{journal}{Phys. Rev. B} \textbf{\bibinfo{volume}{80}},
  \bibinfo{pages}{024519} (\bibinfo{year}{2009}).

\bibitem[{\citenamefont{Petrovic et~al.}(2001)\citenamefont{Petrovic,
  Movshovich, Jaime, Pagliuso, Hundley, Sarrao, Fisk, and
  Thompson}}]{Petrovic2001}
\bibinfo{author}{\bibfnamefont{C.}~\bibnamefont{Petrovic}},
  \bibinfo{author}{\bibfnamefont{R.}~\bibnamefont{Movshovich}},
  \bibinfo{author}{\bibfnamefont{M.}~\bibnamefont{Jaime}},
  \bibinfo{author}{\bibfnamefont{P.~G.} \bibnamefont{Pagliuso}},
  \bibinfo{author}{\bibfnamefont{M.~F.} \bibnamefont{Hundley}},
  \bibinfo{author}{\bibfnamefont{J.~L.} \bibnamefont{Sarrao}},
  \bibinfo{author}{\bibfnamefont{Z.}~\bibnamefont{Fisk}}, \bibnamefont{and}
  \bibinfo{author}{\bibfnamefont{J.~D.} \bibnamefont{Thompson}},
  \bibinfo{journal}{Europhys. Lett.} \textbf{\bibinfo{volume}{53}},
  \bibinfo{pages}{354} (\bibinfo{year}{2001}).

\bibitem[{\citenamefont{Chen et~al.}(2006)\citenamefont{Chen, Matsubayashi,
  Ban, Deguchi, and Sato}}]{Chen2006}
\bibinfo{author}{\bibfnamefont{G.~F.} \bibnamefont{Chen}},
  \bibinfo{author}{\bibfnamefont{K.}~\bibnamefont{Matsubayashi}},
  \bibinfo{author}{\bibfnamefont{S.}~\bibnamefont{Ban}},
  \bibinfo{author}{\bibfnamefont{K.}~\bibnamefont{Deguchi}}, \bibnamefont{and}
  \bibinfo{author}{\bibfnamefont{N.~K.} \bibnamefont{Sato}},
  \bibinfo{journal}{Phys. Rev. Lett.} \textbf{\bibinfo{volume}{97}},
  \bibinfo{pages}{017005} (\bibinfo{year}{2006}).

\end{thebibliography}


 
\onecolumngrid
\clearpage

\appendix
\renewcommand{\thefigure}{S\arabic{figure}} 
\setcounter{figure}{0}
\section{Supplemental Material}

\begin{figure}[!h]
        \includegraphics[width=11cm]{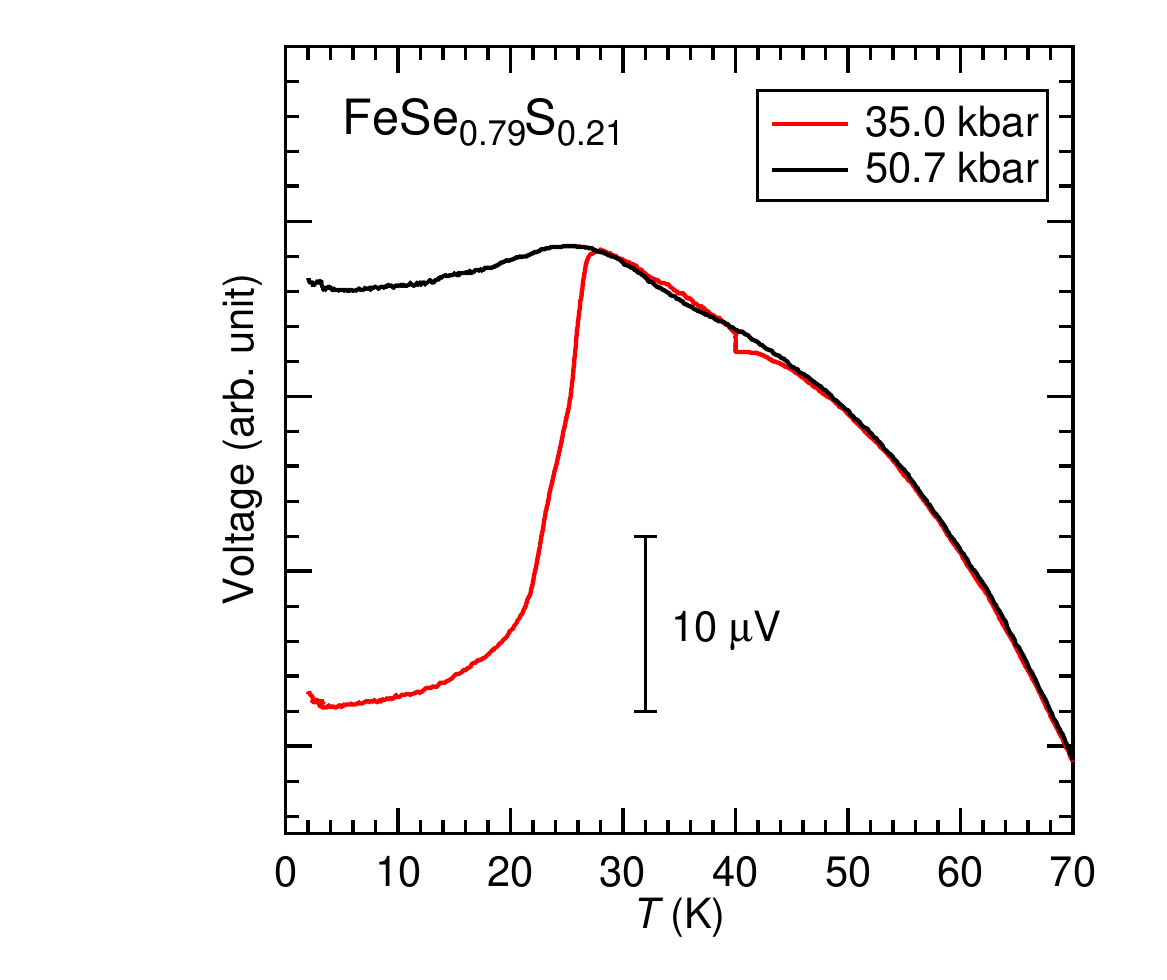}       				
  \caption{\label{figS3}
(Color online) Plot of AC susceptibility against temperature for FeSe$_{1-x}$S$_x$ with $x=0.21$. The background signal is not removed. Voltage drop due to the superconducting transition can be unambiguously identified at 35.0~kbar. The $T_c$ value for this run is included in $T$-$p$ phase diagrams shown in the main text (Figs. 2(b) and 4(c)) as the solid square. }
\end{figure}

One of the AC susceptibility runs of FeSe$_{1-x}$S$_x$ with $x=0.21$ is shown in Fig. \ref{figS3}. For this particular run, a piece of $x=0.21$ was included in the sample space without Y-BSCCO. At 35.0~kbar, a sharp drop in voltage corresponding to the superconducting transition can be seen, even without removing the background signal. At 50.7~kbar, such a substantial voltage drop is not observed. 

\begin{figure}[!h]
        \includegraphics[width=10cm]{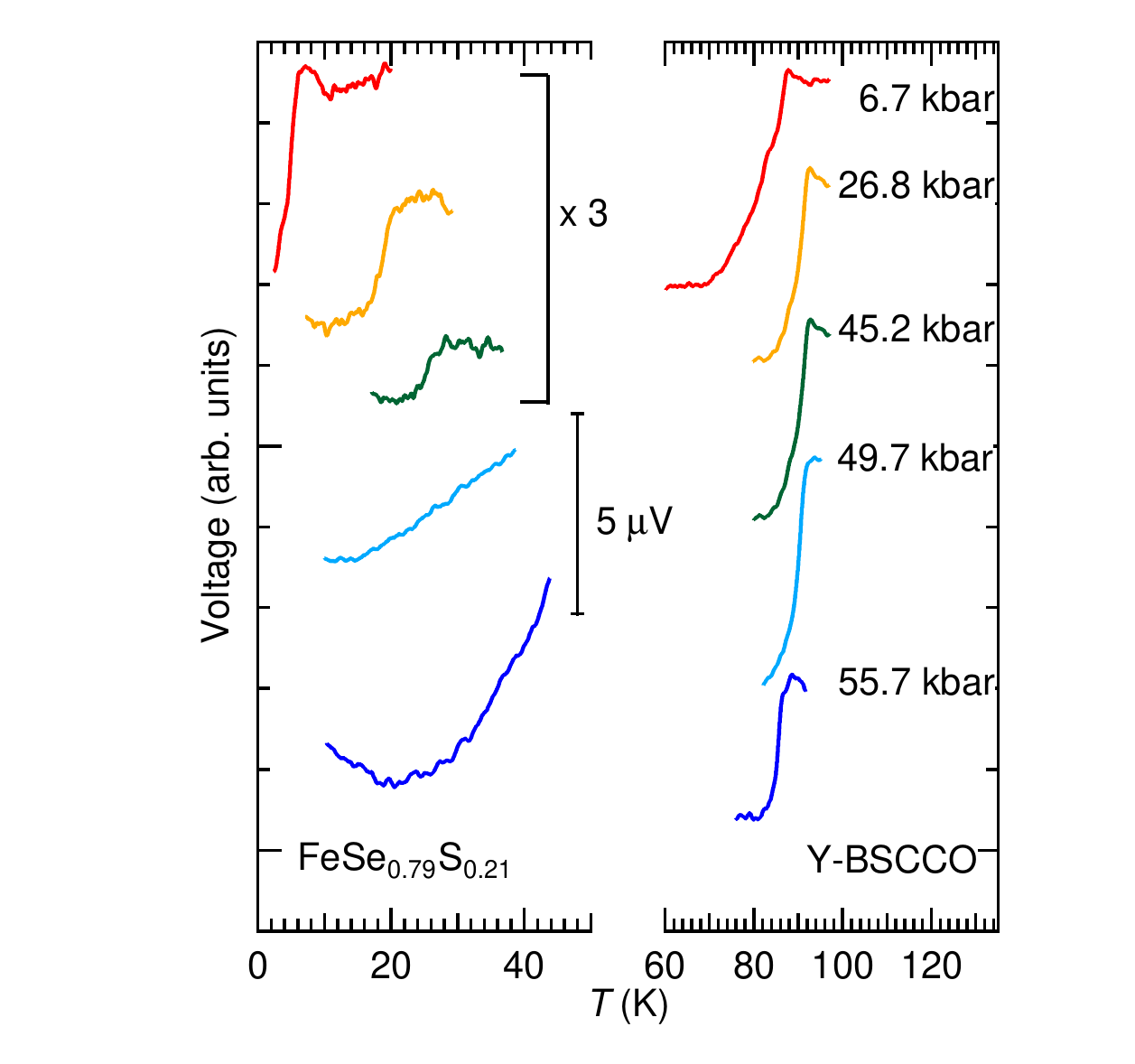}       				
  \caption{\label{figS1}
(Color online) Simultaneous AC susceptibility measurement of Y-BSCCO and FeSe$_{1-x}$S$_x$ ($x=0.21$). The measurements were performed in the order of increasing pressure in a single series without changing the sample. The temperature sweeps are offset vertically for clarity. The axis breaks are added to enable a clear display of both the high-temperature and the low-temperature parts, where the superconducting transitions of Y-BSCCO and FeSe$_{1-x}$S$_x$ occur, respectively. The pressure values are shown next to the high-temperature part only. }
\end{figure}

For completeness, we provide the data from the simultaneous measurement of Y-BSCCO and FeSe$_{1-x}$S$_x$ ($x=0.21$) in Fig. \ref{figS1}. The $T_c$ of Y-BSCCO in this run is slightly lower compared with the other runs shown in Fig. 3 of the main text. However, the pressure dependence of $T_c$ remains qualitatively similar to the other Y-BSCCO crystals we studied ({\it c.f.} Figs. 3 and 4 of the main text). The diamagnetic voltage drop of Y-BSCCO is broader at 6.7~kbar, but it sharpened up rapidly at higher pressures. Note that the diamagnetic voltage drops for FeSe$_{1-x}$S$_x$ ($x=0.21$) for the first three pressure points have been amplified by a factor of 3 for clarity. At 6.7~kbar, the voltage of $x=0.21$ did not level off at the lowest temperature of the run. Therefore, the diamagnetic voltage drop for $x=0.21$ at this pressure point is a lower bound of the estimate.

\begin{figure}[!h]
        \includegraphics[width=10cm]{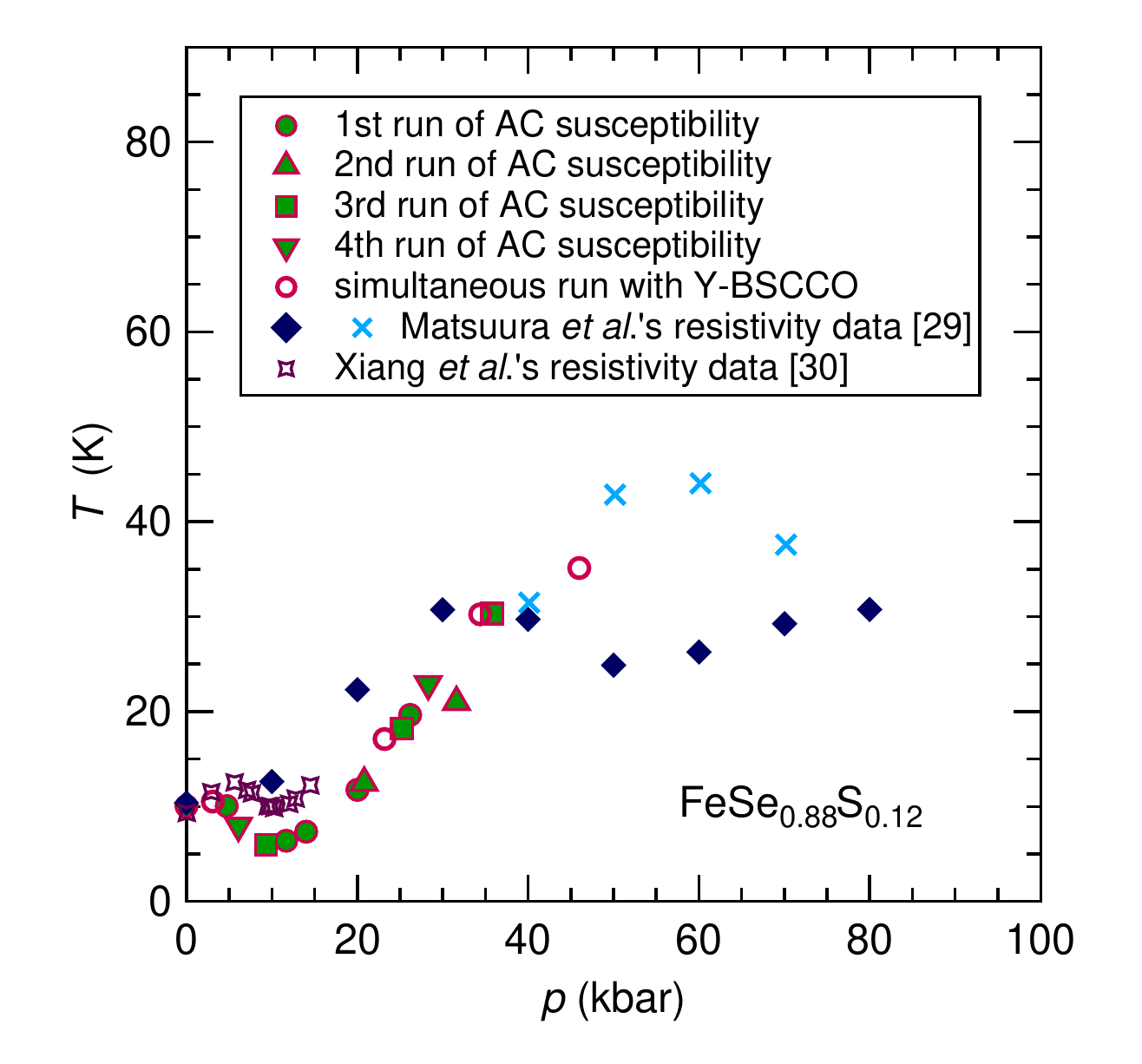}       				
  \caption{\label{figS2}
(Color online) Temperature-pressure phase diagram with data from Matsuura {\it et al.} [29] and Xiang {\it et al.} [30] included. The SDW transition temperatures are denoted by crosses. Other symbols denote $T_c$.}
\end{figure}

In Fig. \ref{figS2}, we compare our $T_c(p)$ data in $x=0.12$ determined with AC susceptibility with that of Matsuura {\it et al.} [29] and Xiang {\it et al.} [30], both determined using resistivity. The comparison with the data of Matsuura {\it et al.} is particularly important because the crystals used in our study are from the same batch as theirs. The SDW transition temperatures, determined resistively, are shown as crosses. The disappearance of our AC susceptibility signals clearly occurs near the verge of magnetism, as discussed in the main text, while resistively-determined $T_c$ persists up to at least 80~kbar.

\end{document}